\documentclass[12pt]{article}
\usepackage{epsf}
\begin{document}
\baselineskip 24pt
\newcommand{\sheptitle}
{Effects of the scale-dependent vacuum expectation values in 
the renormalisation group analysis of neutrino masses}
\newcommand{\shepauthor}
{ N.Nimai Singh\footnote{Permanent Address: Department of Physics,
Gauhati University, Guwahati -781014, India}}

\newcommand{\shepaddress}
{ Department of Physics and Astronomy,
University of Southampton,  Southampton, 
SO17  1BJ,   U.K.}

\newcommand{\shepabstract}
 {The contribution of scale-dependent vacuum expectation values (VEVs)
 of Higgs scalars, which gives significant effects in the evolution of 
fundamental fermion masses in the Minimal Supersymmetric Standard Model 
(MSSM), is now considered  in the derivation of 
the  one-loop analytic expression for  the evolution of the  
left-handed Majorana neutrino masses with energies. The inclusion of such 
 effect of the running VEV 
increases  the stability of the neutrino 
masses under quantum corrections even for low values of 
$\tan\beta \geq 1.42$ at 
the scale $\mu =10^{12}$GeV, and leads to  a mild 
decrease of neutrino masses with higher energies. Such trend 
is common to  that of   other fundamental fermion masses.}

\begin{titlepage}
\begin{flushright}
hep-ph/0009211\\
\end{flushright}
\begin{center}
{\large{\bf \sheptitle}}
\bigskip \\  \shepauthor \\ \mbox{} \\ {\it  \shepaddress} \\ \vspace{.5in}
{\bf Abstract} \bigskip \end{center} \setcounter{page}{0}
\shepabstract
\end{titlepage}

In recent years a large number of theoretical papers were devoted to building
 models for generating small neutrino masses and lepton mixings within or 
outside 
the framework of the Grand Unified Theories (GUTs) with extended $U(1)$ group[1]. 
Both analytic and numerical studies[2,3,4]  have been carried out for 
checking the 
stability of the textures of neutrino mass matrix and lepton mixing matrix  
under quantum 
radiative corrections[5]. There are basically two approaches: the top-down 
approach[2] which predicts the neutrino masses and mixings in terms of 
GUT-parameters, and the bottom-up approach [6] which predicts the running
 parameters at higher scales in terms of experimentally determined values
at low energies. In the top-down programme, one usually 
starts with the running of a set of the RGEs for Yukawa matrices and gauge 
couplings in the MSSM (or SM), with three right-handed heavy neutrinos, 
taking into account the effects of the heavy neutrino mass thresholds, 
from GUT 
scale down to the lightest right-handed neutrino mass scale $(M_{R1})$. 
This fixes the left-handed Majorana neutrino mass matrix $m_{LL}(M_{R1})$ 
through the see-saw mechanism[7] at this scale,
\begin{equation}
m_{LL}(M_{R1})=v^2_{u}Y_{\nu}(M_{R1})M^{-1}_{RR}Y_{\nu}^{T}(M_{R1})
\end{equation}
Below this scale $M_{R1}$ the right-handed neutrinos decouple from the theory,
and the neutrino mass matrix in Eq.(1) is taken as[2] 
\begin{equation}
m_{LL}(M_{R1})=v^2_{u}\kappa (M_{R1})
\end{equation}
where $\kappa$ is the coefficient of the dimension 5 neutrino mass operator.
In the energy range from $M_{R1}$ down to 
low energy at $m_t$, the running of the coefficient $\kappa$ in the diagonal
 charged lepton basis, 
fixes the neutrino mass matrix at scale $m_t$,
\begin{equation}
m_{LL}(m_t)=v_u^2\kappa(m_t).
\end{equation}
 In the above discussion only the scale-dependence of 
$\kappa$ is considered, and not the running of  the vacuum
expectation value(VEV),$v_u$ in Eqs.(1)-(3). This led  to the increase of 
neutrino mass eigenvalues with  energy scales, giving  significant effect 
 for  low $\tan\beta$ values. 
As it is strongly $\tan\beta$ dependent, this efect may lead to  the 
instability of the neutrino masses 
under quantum radiative corrections\footnote
{In Refs.[3,4] the stability condition is decided by the change in 
texture of neutrino mass matrix only. Here we emphasise that changing 
pattern of the overall magnitudes of neutrino mass eigenvalues at different 
energies, may also cause instability.}. For higher values of $\tan\beta$
the stability is again improved. 
Such  increasing trend of neutrino mass eigenvalues 
with the increase in  energies,  is opposite to that of the  general trend 
shown by 
other fundamental fermions (charged leptons and quarks)[8,9].
The effects of the contributions of the scale-dependent vacuum expectation values(VEVs)
of Higgs scalars in the one-loop analytic expressions, in the evolution of 
quarks 
and charged leptons masses at higher energies in the MSSM, had been studied
in Ref.(8), and this effect is quite significant.

  In this paper we study the stability of the magnitudes of neutrino 
masses at low $\tan\beta$ and their running behaviour at different energies,
 by considering
  the scale-dependence[10] of the vacuum expectation 
value (VEV), $v_{u}$, along with that of $\kappa$. The expression in Eq.(3)
 is now modified as  
\begin{equation}
m_{LL}(t)=v_u^2(t)\kappa(t)
\end{equation}
where $v_u(t_0)=v_0\sin\beta$,  $v_0=174$GeV,  $t=\ln\mu, 
t_0=\ln m_t$.  The above equation(4) can be written as 
\begin{equation}
\frac{d\ln m_{LL}(t)}{dt}=\frac{d\ln\kappa (t)}{dt} +
      2\frac{d\ln v_{u}(t)}{dt} 
\end{equation}
where the second term on the right-hand side of the above equation 
is the contribution from the running of the VEV. 
The RGEs for $v_u$ [8,10] and $\kappa$ [2,5]in the diagonal charged 
lepton basis, for one-loop order in MSSM, in   the energy range 
$t\geq t_0$, are given by 
\begin{equation}
\frac{d\ln v_u}{dt}=\frac{1}{16\pi^2}[\frac{3}{20}g_1^2 + \frac{3}{4}g_2^2
-3h_t^2],
\end{equation}
and,
\begin{equation}
\frac{d\ln\kappa}{dt}=-\frac{1}{16\pi^2}[\frac{6}{5}g_1^2 + 6g_2^2
-6h_t^2 -\delta_{i3}h_\tau^2 - \delta_{3j}h_\tau^{2}]
\end{equation}
respectively. Substituition of  Eqs.(6,7) in Eq.(5) gives
\begin{equation}
\frac{d\ln m_{LL}}{dt}=\frac{1}{16\pi^2}[-\frac{9}{10}g_1^2 - \frac{9}{2}g_2^2
 +\delta_{i3}h_\tau^2 + \delta_{3j}h_\tau^{2}]
\end{equation}
Upon integration from low scale $t_0=\ln m_t$ to high scale $t_{R1}=\ln M_{R1}$ where 
$t_{R1}\geq t_0$,  
we get the correct expression for the neutrino mass matrix at $t_0$,
\begin{equation} 
\frac{(m_{LL}(t_0))_{ij}}{(m_{LL}(t_{R1}))_{ij}}=e^{(\frac{9}{10}I_{g1} +
 \frac{9}{2}I_{g2})}
 e^{-I_{\tau}(\delta_{i3} + \delta_{3j})},
\end{equation}
\begin{equation}
I_f=\frac{1}{16\pi^2}\int_{\ln m_t}^{\ln M_{R1}}h_{f}^{2}(t)dt,
\end{equation}
\begin{equation}
I_{g_i}=\frac{1}{16\pi^2}\int_{\ln m_t}^{\ln M_{R1}}g_{i}^2(t)dt
\simeq\ln\left(\frac{g_{i}(t_{R1})}{g_{i}(t_0)}\right)^{(1/b_i)}
\end{equation}
where $f=t,\tau$; $i=1,2,3$, and $b_i=(33/5,1,-3)$ for MSSM. The correct 
expression in Eq.(9) will
 certainly affect the earlier numerical results obtained without taking 
the effect of the running VEV [2]
 at scale $M_{R1}$.

For simplicity we now follow the analysis of  the RGEs for neutrino mass 
eigenvalues[4].
With the inclusion of such scale-dependence of VEV in  Eq.(6), 
the RGEs for 
 the mass eigenvalues given in Ref.[4] in the diagonal charged lepton 
basis, is now modified as,
\begin{equation}
\frac{d\ln m_{\nu a}}{dt}=\frac{1}{16\pi^2}\sum_{b=e,\mu,\tau}[
-\frac{9}{10}g_1^2 - \frac{9}{2}g_2^2 + 2h_b^{2}V_{ba}^2]
\end{equation}
where $a=1,2,3$, and $V_{ba}$ is the MNS mixing matrix element. 
The correct expression for the neutrino mass ratio at different 
energy scales, is also 
obtained by integrating Eq.(12) as 
\begin{equation} 
R_{a}(t_{R1})=\frac{m_{\nu a}(t_{R1})}{m_{\nu a}(t_{0})}\approx 
e^{-(\frac{9}{10}I_{g1}+ \frac{9}{2}I_{g2})}
 e^{2 V_{\tau a}^{2}I_{\tau}}
\end{equation}
In  getting Eq.(13) we have neglected very small effects due to $I_{\mu,e}$
compared to $I_{\tau}$, and also assumed $V_{\tau a}$ does not change much 
in the integration range\footnote { Such approximation can  be  justified
 for the calculation of the mass eigenvalues and their ratios as the 
second exponential term in Eq.(13) gives  almost 1 for low values of 
$I_{\tau}$.}.
For a typical value of the element of MNS mixing matrix 
$V_{\tau3}\simeq \frac{1}{\sqrt{2}}$, 
we can get the condition $m_{\nu3}(t_0) >  m_{\nu3}(t_{R1})$ following 
Eq.(13), which shows 
a mild increase in neutrino masses  with the decrease  in energies,
 even for small $\tan\beta\geq 1.42$. This is due to the fact that the 
ratio $R_{3}(t_{R1})$
is now independent of  $e^{6I_t}$  in the first exponential factor in Eq.(13).
 The same is true in Eq.(9). In fact the contribution of the running VEV
effectively makes the replacement in the exponential factor:
\begin{equation}
e^{-(\frac{6}{5}I_{g1} + 6I_{g2}-6I_t)}\rightarrow 
 e^{-(\frac{9}{10}I_{g1} + \frac{9}{2}I_{g2})}
\end{equation}
 in Eqs.(9,13).

We now study the effect of the running VEV in the evolution of squared 
neutrino 
mass difference, 
$\bigtriangleup m^{2}_{ij}= |m_{\nu i}^{2}-m_{\nu j}^{2}|$ 
with energies. By taking square on both sides of   Eq.(13), and considering 
for two mass eigenvalues $a=i,j$,  we get approximately,
\begin{equation} 
\bigtriangleup m_{ij}^{2}(t_{R1})\approx 
\bigtriangleup m_{ij}^{2}(t_{0})e^{-2(\frac{9}{10}I_{g1}
 + \frac{9}{2}I_{g2})}
 e^{4 V_{\tau i}^{2}I_{\tau}}
\end{equation}
where we assume that the small  difference between $V_{\tau i}$ and
 $V_{\tau j}$ for $i,j=1,2,3$, does not alter much the last exponential term
which can be approximately taken as $e^{4V^{2}_{\tau i}I_{\tau}}\simeq 
e^{4V^{2}_{\tau j}I_{\tau}}\approx 1$ for the low values  of $I_{\tau}$. 
This amounts to neglecting small changes in the texture of neutrino 
mass matrix which would be relevant for the evolution of mixing angles.
 The evolution of $\bigtriangleup m_{ij}^{2}(t_{R1})$
 is now stable with the effects of running VEV for both low and high values
 of $\tan\beta$, otherwise it would have been 
more strongly $\tan\beta$-dependent with $e^{12I_{t}}$ in the exponential 
factor in the case where  the effect of running VEV is not included,
 causing more instability at low $\tan\beta$ values.

The running of  the   ratio of two neutrino mass eigenvalues,
$R_{23}=m_{\nu2}/m_{\nu3}$ 
(and hence the running of $RR_{23}$) 
is independent of the effect of running VEV, so that the ratio of ratios is,
\begin{equation} 
RR_{23}(t_{R1})=\frac{R_{23}(t_{R1})}{R_{23}(t_{0})}\approx e^{-2\delta V_{\tau 32}^{2}I_{\tau}}
\end{equation}
where 
\begin{equation} 
\delta V_{\tau 32}^{2}=V_{\tau 3}^{2} - V_{\tau 2}^{2}
\end{equation}
which can be either  positive,  negative or zero.  For the  positive value, 
$\delta V_{\tau 32}^{2}>0$
as in hierarchical 
case[2], one gets  the condition, 
\begin{equation}
R_{23}(t_0)\geq R_{23}(t_{R1})
\end{equation}
which implies the increase in  the neutrino mass ratio $m_{\nu2}/m_{\nu3}$ 
 with the decrease in energies[2]. If we start with degenerate 
neutrinos,
$m_{\nu2}=m_{\nu3}$ at the scale $M_{R1}$, then we would get 
$m_{\nu2}>m_{\nu3}$ at scale $m_t$. This shows that nearly degenerate 
neutrinos are not stable under quantum corrections[3].

The above relations in Eqs.(16-18) for $a=2,3$, can be generalised 
for any  pair of mass eigenvalues $a=i,j$. For inverted hierarchical case [2] 
with $m_{\nu 1}>m_{\nu 2}$, we may have $\delta V^{2}_{\tau 21}<0$
which leads to 
\begin{equation}
R_{12}(t_0)\leq R_{12}(t_{R1})
\end{equation}
where the neutrino mass ratio $m_{\nu 1}/m_{\nu 2}$ decrease with 
the decrease in energies[2]. The effect of the running VEV does not 
change the textures of neutrino mass matrix and hence the MNS mixing matrix.

  Next we turn to numerical analysis of the RGEs in the bottom-up approach
 in running from low energy scale $t_0$ to high energy scale, 
replacing $t_{R1}$ by 
running t in the above equations (9-19). 
We make use of the following input values of the running fermion masses
$m_{i}(m_i)$ of the third family:
\begin{equation}
m_{t,b,\tau}=(166.5, 4.2, 1.785)GeV 
\end{equation}
where, for heavy flavours(top and bottom quarks) the values are derived from
 input pole-masses $m_t^{pole}=175.6$GeV [11] and $m_b^{pole}=4.7$GeV [12,13]
using 
two-loop RGEs in QCD. The initial input values for  the top, the bottom and 
$\tau$-lepton Yukawa couplings at top-quark mass scale $t_0=\ln m_t$ in the RGEs in MSSM, are usually obtained as
\begin{equation}
h_{t}(t_0)=m_t/(174\sin\beta),  
h_{b,\tau}(t_0)=m_{b,\tau}/(174\eta_{b,\tau}\cos\beta).
\end{equation}
Using the CERN-LEP measurements at $M_{Z}=91.18$GeV,
\begin{equation}
\alpha_{3}(M_Z)=0.118\pm 0.004, \alpha^{-1}(M_Z)=127.9\pm0.1,
 \sin^{2}\theta_{\omega}(M_Z)=0.2313\pm 0.0003,
\end{equation}
we obtain the values of gauge couplings at scale $t_0$ using one-loop RGEs,
assuming the existence of one-light Higgs doublet (n=1) and five quark flavours below $m_t$ scale,
\begin{equation}
\alpha^{-1}_{1,2,3}(t_0)=58.42,29.67,8.89
\end{equation}
The QCD-QED rescaling factors [6] are calculated as 
\begin{equation}
\eta_{f}=(1.54,1.017),f=b,\tau
\end{equation}
As a result of the numerical analysis of the RGEs for Yukawa and gauge
couplings at two-loop level[6] in the energy range $t_0<t<t_{U}$,
 the unification of three 
gauge couplings  is observed at $M_{U}=1.82\times10^{16}$GeV. 
The values of Yukawa couplings $(h_t,h_b,h_{\tau})$, gauge couplings 
and values of integrals $I_{i}$ defined in Eqs.(10,11) for different
 values of  $\tan\beta=1.42-60.0$
 are estimated at different energy scales.

We present our numerical results in Figs.(1-4) where the solid line 
refers to the analysis with the effects of running VEV in the present 
calculation (referred to as case A).
We also present the corresponding results  without the  effect of 
running VEV in dotted line (referred to as case B) for comparison only. 
 With a typical  
input value  $V_{\tau 3}=\frac{1}{\sqrt{2}}$, the variation of the ratio 
$R_{3}(t)$ defined in Eq.(13), with energy scales  t for two 
representative values of  
$\tan\beta=1.63$ and
 $ 57.29$, are presented in Figs.1 and 2 respectively. These figures show the 
evolutions of neutrino
 mass eigenvalue $m_{\nu 3}$ with the increase in energy scale.

We observe that for high value of $\tan\beta = 57.29$ in Fig.2,
 the evolution of the ratio $R_{3}(t)=m_{\nu3}(t)/m_{\nu3}(t_0)$ is 
almost stable in both cases A and B.  However, for low value of $\tan\beta=1.63$
 in Fig.1, there is a significant  increase in $R_{3}(t)$ at higher energies 
in case B. For example, at $\mu=1.82\times10^{16}$ GeV, the ratio $R_{3}(t)$
  is
about $5.63$ in case B as shown in Fig.1 by dotted line. 
Such  unwanted feature 
which may cause instability, is not present in  
case A  (solid line in Fig.1). Fig.3 shows the variation 
of neutrino mass, $R_{3}(t_{R1})=m_{\nu 3}(t_{R1})/m_{\nu 3}(t_0)$ at a 
particular scale, $t_{R1}=27.63 $ corresponding to $M_{R1}= 10^{12}GeV$,
  with different values of 
$\tan\beta = 1.42 - 60$. We see that 
at low $\tan\beta\geq 1.42$ region there is a significant enhancement in 
 $R_{3}\leq 5.3$ in case B whereas the ratio is stable in case A for all 
values
of $\tan\beta$. 
For higher values of $\tan\beta$ the ratio is again stable in case B. 
The same analysis is true  for the cases of other two mass eigenvalues 
with $a=1,2$.
Similar analysis can be done for the evolution of 
$\bigtriangleup m^{2}_{ij}$ in Eq.(15), which would be  very unstable 
in the  low $\tan\beta$ region in case B. However, it  is now stable 
for all values of $\tan\beta$ under radiative corrections at higher energies in case A.


\vbox{
\noindent
\hfil
\vbox{
\epsfxsize=10cm
\epsffile[130 380 510 735]{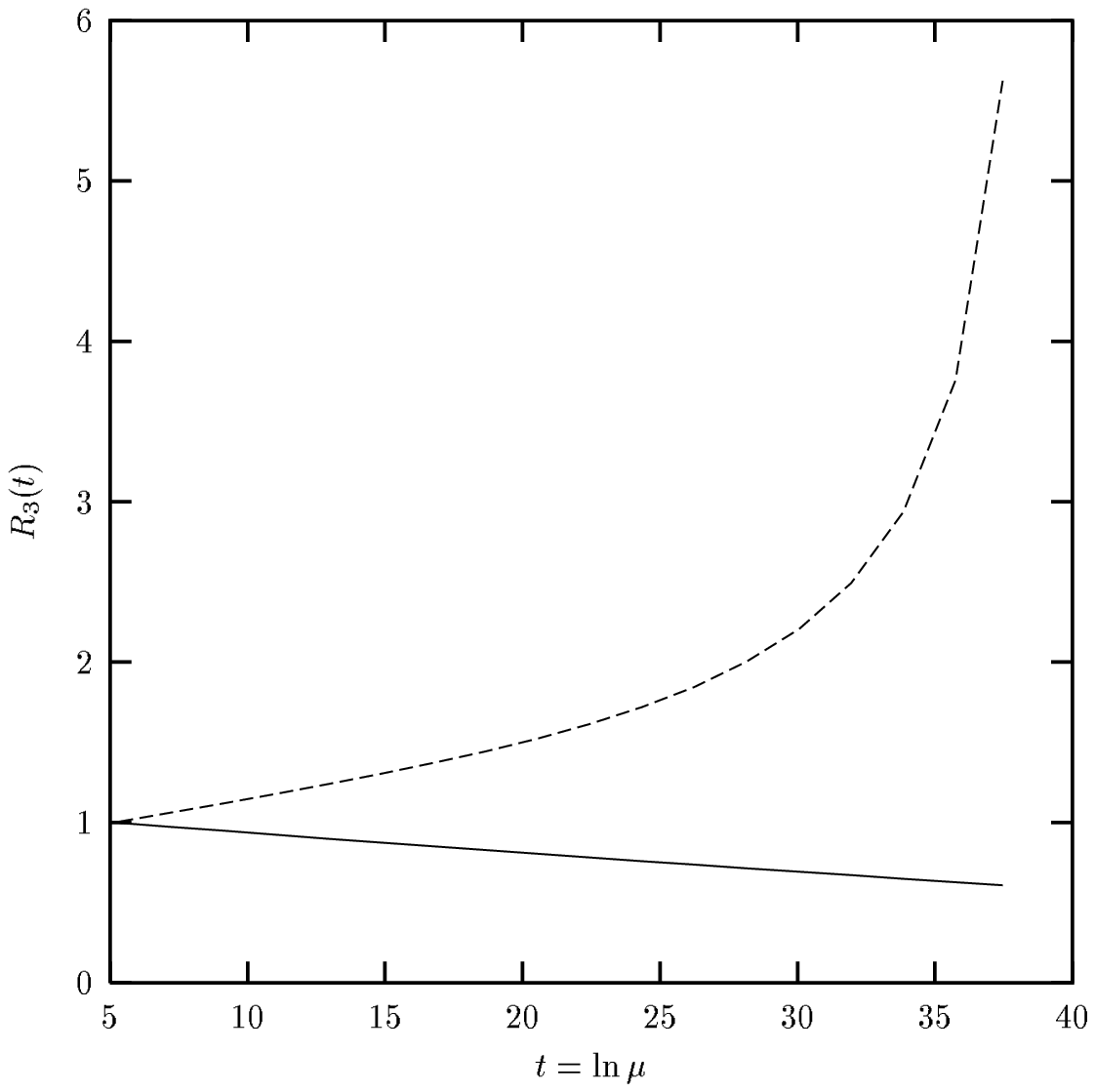}}

{\narrower\narrower\footnotesize\noindent
{Fig.1}
Variation of  $R_{3}(t)=m_{\nu 3}(t)/m_{\nu 3}(t_0)$
with energies $t=\ln\mu$ for small value of $\tan\beta=1.63$. The results
with and without the effect of running VEV, are shown in solid line 
and dotted line  respectively. 
\par}}

\vbox{
\noindent
\hfil
\vbox{
\epsfxsize=10cm
\epsffile[130 380 510 735]{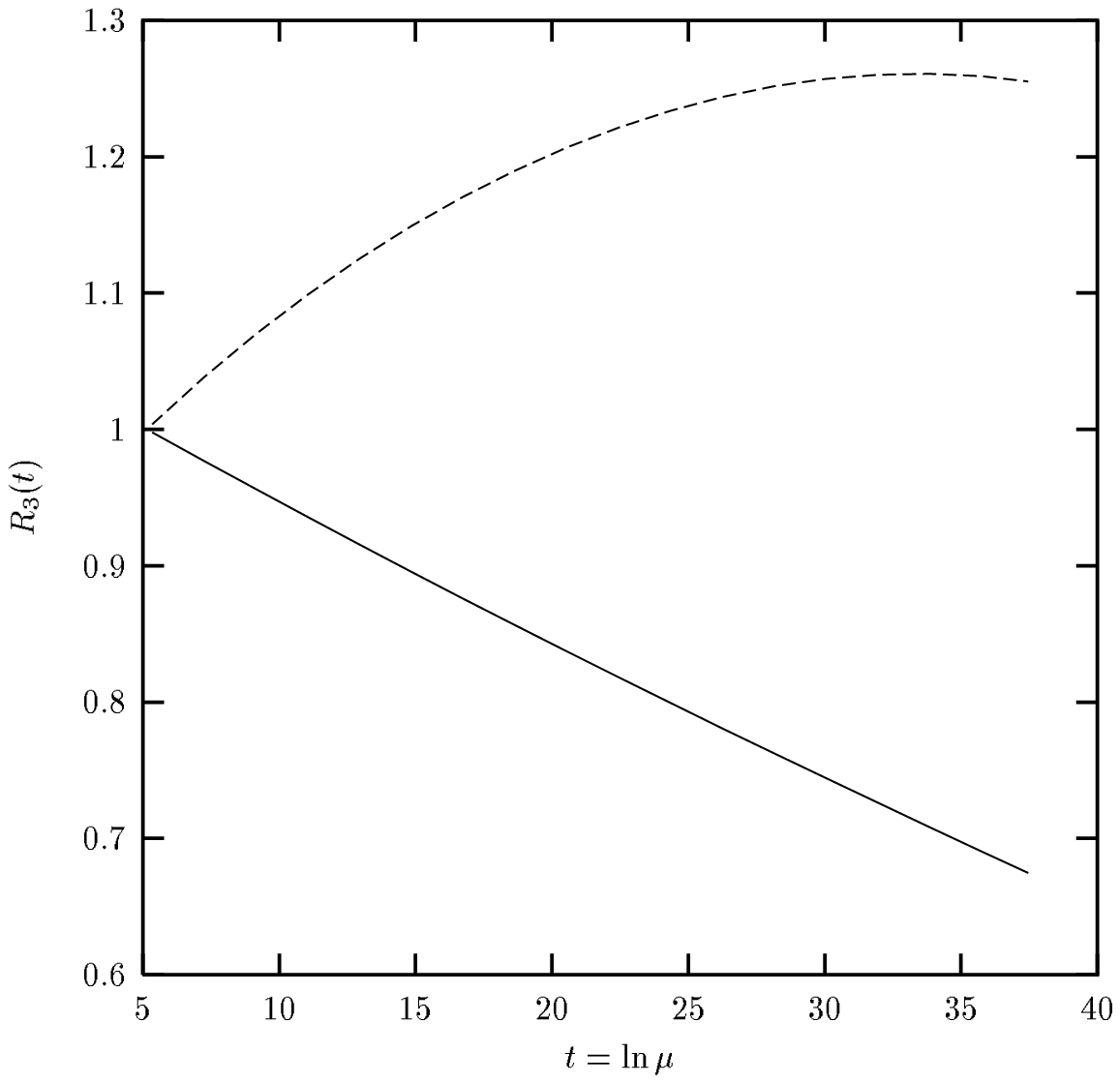}}

{\narrower\narrower\footnotesize\noindent
{Fig.2}
Variation of $R_{3}(t)=m_{\nu 3}(t)/m_{\nu 3}(t_0)$
with energies $t=\ln\mu$ for large  value of $\tan\beta=57.29$. The results
with and without the effect of running VEV are shown in 
solid line  and dotted line  respectively. 

\par}}


\vbox{
\noindent
\hfil
\vbox{
\epsfxsize=10cm
\epsffile[130 380 510 735]{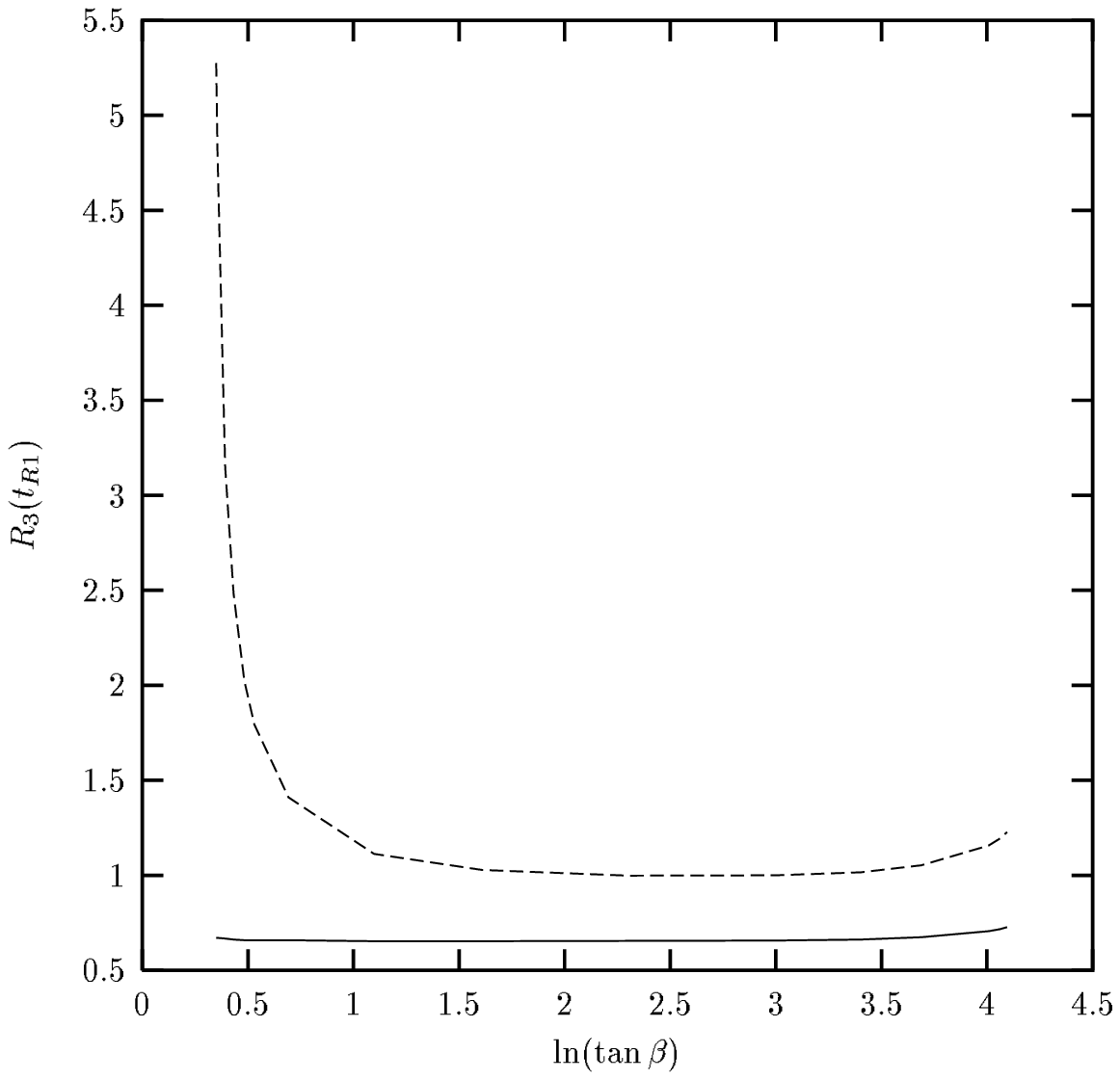}}

{\narrower\narrower\footnotesize\noindent
{Fig.3}
Variation of 
 $R_{3}(t_{R1})=m_{\nu 3}(t_{R1})/m_{\nu 3}(t_0)$
with $\ln(\tan\beta)$ for $M_{R1}=10^{12}GeV$. The results
with and without the effect of running VEV are shown in solid line 
 and dotted line  respectively. 
\par}}

\vbox{
\noindent
\hfil
\vbox{
\epsfxsize=10cm
\epsffile[130 380 510 735]{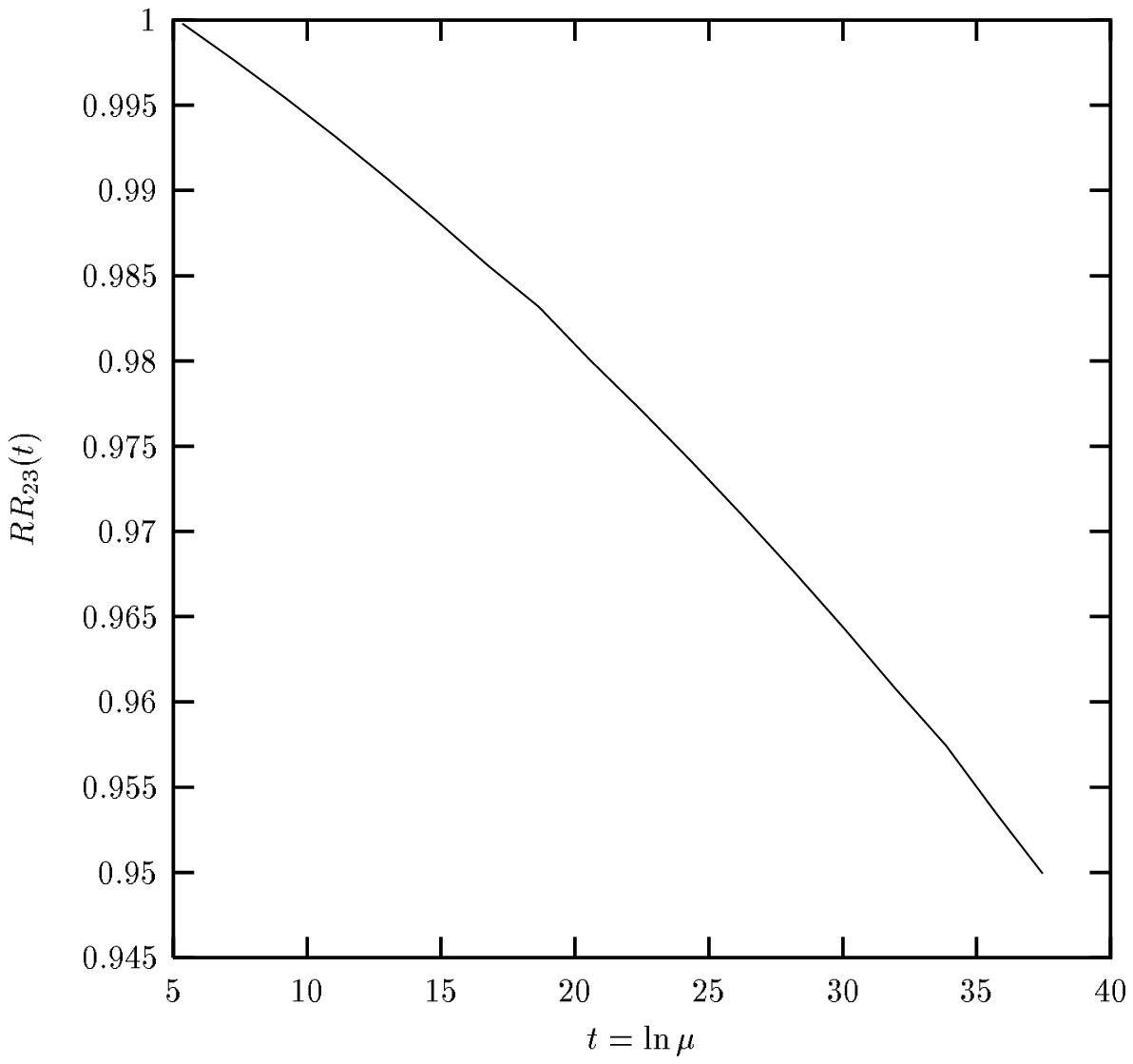}}

{\narrower\narrower\footnotesize\noindent
{Fig.4}
Variation of the ratio of the neutrino mass ratio 
$RR_{23}(t)=(\frac{m_{\nu 2}(t)}
{m_{\nu 3}(t)})/ (\frac{m_{\nu 2}(t_0)}
{m_{\nu 3}(t_0)})$
with energies $t=\ln\mu$ for large  value of $\tan\beta=57.29$. 

\par\bigskip}}




Finally, we study the relative rates of the evolution of two neutrino mass 
eigenvalues in terms of their ratio, $R_{23}=m_{\nu2}/m_{\nu3}$  
given in Eq.(16), in going  from low to high 
energies. We consider  high value of $\tan\beta=57.29$ where the effect of 
$I_{\tau}$ is large, and this ratio increases with the decrease in
energies by a few percent only. This is shown in Fig.4 where we present 
the evolution of the ratio of the ratios $RR_{23}(t)$ in Eq.(16) 
with energies. 
This leads to a mild increase in the  hierarchical relation, 
$m_{\nu 2}/m_{\nu 3}$ at lower energies. As noted earlier,
 such hierarchical ratios are 
independent of the effect of running VEV. 
Finally we point  out  the  changes aring from the  running of VEV in 
 the earlier calculations[2] of  neutrino masses. The earlier results at 
low  scale $m_t$
in Ref.[2] do not change at all. However if  we prefer to express  the 
neutrino mass matrix at higher scale $M_{R1}$, then we have to take the 
effect of running VEV, $v_{u}(t_{R1})$ in place of $v_{u}(t_0)$,
 which modifies the earlier numerical results at 
the scale $M_{R1}$.

To conclude, we  have  considered  the contributions of scale-dependent 
vacuum expectation values (VEVs) of Higgs scalars in deriving one-loop 
analytic expression for running the 
left-handed Majorana neutrino masses with energies 
in the MSSM. This gives significant changes in the expression of the 
evolution of neutrino masses, and  also increases  the stability 
of the neutrino 
masses under quantum corrections even for low $\tan\beta$. We observed 
 a mild decreasing trend of neutrino masses with higher energies, which  
is now common to that of all  other fermion masses in nature.

\section*{Acknowledgement}
The author would like to thank S.F.King for useful discussions and M.Oliveira
 for helping in computations.

\end{document}